\documentclass[10pt]{article}
\usepackage[fleqn]{amsmath}
\usepackage{amsmath,amsthm,amsfonts,comment,array,braket}

\DeclareMathOperator{\Hom}{Hom}
\DeclareMathOperator{\Com}{Com}

\DeclareMathOperator{\1}{id}

\newcommand{\NN}{\mathbb{N}}
\newcommand{\RR}{\mathbb{R}}

\newcommand{\EEnd}{\mathcal End}
\newcommand{\EE}{\mathcal E}

\renewcommand{\=}{:=}
\renewcommand{\t}{\otimes}
\renewcommand{\:}{\colon}

\newcommand{\la}{\lambda }
\newcommand{\m}{\overset{\circ}{\mu}}
\newcommand{\A}{\hat{A}}
\newcommand{\pp}{\hat{p}}
\newcommand{\q}{\hat{q}}
\newcommand{\muu}{\hat{\mu}}
\newcommand{\qxi}{\hat{\xi}}

\newcommand{\ee}{\varepsilon}
\renewcommand{\H}{\hat{H}}
\newcommand{\e}{\hat{\varepsilon}}
\newcommand{\pn}{\sqrt{2E_n}}

\newtheorem{thm}{Theorem}[section]

 \newtheorem{lemma}[thm]{Lemma}
 \newtheorem{cor}[thm]{Corollary}

\theoremstyle{definition}
 \newtheorem{defn}[thm]{Definition}

\theoremstyle{definition}

\theoremstyle{definition}

\numberwithin{equation}{section}
\numberwithin{table}{section}

\allowdisplaybreaks

\begin{document}
\title{\LARGE\bf VII$^{\hbar}_a$, III$_{a=1}^{\hbar}$, VI$_{a\neq1}^{\hbar}$}
\author{\large Eugen Paal and J\"{u}ri Virkepu
}
\date{}
\maketitle
\thispagestyle{empty}
\begin{abstract}
Operadic Lax representations for the harmonic oscillator are used to construct the quantum counterparts of some 3d real Lie algebras in Bianchi classification. The Jacobians of these quantum algebras are studied. It is conjectured that the tangent algebras of these quantum algebras are the Heisenberg algebra. From this it follows that the volume element  in $\RR^{3}$ is quantized by $|(x,y,z)|=4\sqrt{2}(2n+1)$, ($n=0,1,2,\dots$). Thus, the elementary (minimal) length in this model is $l_{min}=2^{5/6}$.
\end{abstract}

\section{Introduction}

In Hamiltonian formalism, a mechanical system is described by the
canonical variables $q^i,p_i$ and their time evolution is prescribed
by the Hamiltonian equations
\begin{equation}
\label{ham} \dfrac{dq^i}{dt}=\dfrac{\partial H}{\partial p_i}, \quad
\dfrac{dp_i}{dt}=-\dfrac{\partial H}{\partial q^i}
\end{equation}
By a Lax representation \cite{Lax68,BBT03} of a mechanical system
one means such a pair $(L,M)$ of matrices (linear operators) $L,M$
that the above Hamiltonian system may be represented as the Lax
equation
\begin{equation}
\label{lax} \dfrac{dL}{dt}= ML-LM
\end{equation}
Thus, from the algebraic point of view, mechanical systems can be
represented by linear operators, i.e by  linear maps $V\to V$  of a
vector space $V$. As a generalization of this one can pose the
following question \cite{Paal07}: how can the time evolution
of the linear operations (multiplications) $V^{\t n}\to V$ be described?

The algebraic operations (multiplications) can be seen as an example
of the \emph{operadic} variables \cite{Ger}. If an operadic system
depends on time one can speak about \emph{operadic dynamics}
\cite{Paal07}. The latter may be introduced by simple and natural
analogy with the Hamiltonian dynamics. In particular, the time
evolution of the operadic variables may be given by the operadic Lax
equation. In \cite{PV07,PV08,PV08-1}, the low-dimensional binary operadic Lax
representations for the harmonic oscillator were constructed.

In the present paper, the operadic Lax representations for the harmonic oscillator are used to construct the quantum counterparts of some 3d real Lie algebras in Bianchi classification. The Jacobians of these quantum algebras are studied. It is conjectured that the tangent algebras of these quantum algebras are the Heisenberg algebra. From this it follows that the volume element  in $\RR^{3}$ has discrete values: $|(x,y,z)|=4\sqrt{2}(2n+1)$, ($n=0,1,2,\dots$). Thus, the elementary (minimal) length in this model is $l_{min}=2^{5/6}$.

\section{Endomorphism operad and Gerstenhaber brackets}

Let $K$ be a unital associative commutative ring, $V$ be a unital
$K$-module, and $\EE_V^n\= {\EEnd}_V^n\= \Hom(V^{\t n},V)$
($n\in\NN$). For an \emph{operation} $f\in\EE^n_V$, we refer to $n$
as the \emph{degree} of $f$ and often write (when it does not cause
confusion) $f$ instead of $\deg f$. For example, $(-1)^f\= (-1)^n$,
$\EE^f_V\=\EE^n_V$ and $\circ_f\= \circ_n$. Also, it is convenient
to use the \emph{reduced} degree $|f|\= n-1$. Throughout this paper,
we assume that $\t\= \t_K$.

\begin{defn}[endomorphism operad \cite{Ger}]
\label{HG} For $f\t g\in\EE_V^f\t\EE_V^g$ define the \emph{partial
compositions}
\[
f\circ_i g\= (-1)^{i|g|}f\circ(\1_V^{\t i}\t g\t\1_V^{\t(|f|-i)})
\quad \in\EE^{f+|g|}_V,
         \quad 0\leq i\leq |f|
\]
The sequence $\EE_V\= \{\EE_V^n\}_{n\in\NN}$, equipped with the
partial compositions $\circ_i$, is called the \emph{endomorphism
operad} of $V$.
\end{defn}

\begin{defn}[total composition \cite{Ger}]
The \emph{total composition}
$\circ\:\EE^f_V\t\EE^g_V\to\EE^{f+|g|}_V$ is defined by
\[
f\circ g\= \sum_{i=0}^{|f|}f\circ_i g\quad \in \EE_V^{f+|g|}, \quad |\circ=0
\]
The pair $\Com\EE_V\= \{\EE_V,\circ\}$ is called the \emph{composition
algebra} of $\EE_V$.
\end{defn}

\begin{defn}[Gerstenhaber brackets \cite{Ger}]
The  \emph{Gerstenhaber brackets} $[\cdot,\cdot]$ are defined in
$\Com\EE_V$ as a graded commutator by
\[
[f,g]\= f\circ g-(-1)^{|f||g|}g\circ f=-(-1)^{|f||g|}[g,f],\quad
|[\cdot,\cdot]|=0
\]
\end{defn}

The \emph{commutator algebra} of $\Com \EE_V$ is denoted as
$\Com^{-}\!\EE_V\= \{\EE_V,[\cdot,\cdot]\}$. One can prove (e.g
\cite{Ger}) that $\Com^-\!\EE_V$ is a \emph{graded Lie algebra}. The
Jacobi identity reads
\[
(-1)^{|f||h|}[[f,g],h]+(-1)^{|g||f|}[[g,h],f]+(-1)^{|h||g|}[[h,f],g]=0
\]

\section{Operadic Lax equation and harmonic oscillator}

Assume that $K\= \RR$ or $K\= \mathbb{C}$ and operations are
differentiable. Dynamics in operadic systems (operadic dynamics) may
be introduced by

\begin{defn}[operadic Lax pair \cite{Paal07}]
Allow a classical dynamical system to be described by the
Hamiltonian system \eqref{ham}. An \emph{operadic Lax pair} is a
pair $(L,M)$ of homogeneous operations $L,M\in\EE_V$, such that the
Hamiltonian system  (\ref{ham}) may be represented as the
\emph{operadic Lax equation}
\[
\frac{dL}{dt}=[M,L]\= M\circ L-(-1)^{|M||L|}L\circ M
\]
The pair $(L,M)$ is also called an \emph{operadic Lax representations} of/for Hamiltonian system \eqref{ham}.
Evidently, the degree constraints $|M|=|L|=0$ give rise to ordinary
Lax equation (\ref{lax}) \cite{Lax68,BBT03}. In this paper we assume that $|M|=0$.
\end{defn}

The Hamiltonian of the harmonic oscillator (HO) is
\[
H(q,p)=\frac{1}{2}(p^2+\omega^2q^2)
\]
Thus, the Hamiltonian system of HO reads
\begin{equation}
\label{eq:h-osc} \frac{dq}{dt}=\frac{\partial H}{\partial p}=p,\quad
\frac{dp}{dt}=-\frac{\partial H}{\partial q}=-\omega^2q
\end{equation}
If $\mu$ is a linear algebraic operation we can use the above
Hamilton equations to obtain
\[
\dfrac{d\mu}{dt} =\dfrac{\partial\mu}{\partial
q}\dfrac{dq}{dt}+\dfrac{\partial\mu}{\partial p}\dfrac{dp}{dt}
=p\dfrac{\partial\mu}{\partial
q}-\omega^2q\dfrac{\partial\mu}{\partial p}
 =[M,\mu]
\]
Therefore, we get the following linear partial differential equation
for $\mu(q,p)$:
\begin{equation}
\label{eq:diff}
p\dfrac{\partial\mu}{\partial
q}-\omega^2q\dfrac{\partial\mu}{\partial p}=[M,\mu]
\end{equation}
By integrating \eqref{eq:diff} one can get collections of operations called the
\emph{operadic} (Lax representations of) harmonic oscillator. 

\section{Operadic Lax representations of harmonic oscillator}

\begin{lemma}
\label{lemma:harmonic3} Matrices
\[
L\=\begin{pmatrix}
    p & \omega q & 0 \\
    \omega q & -p & 0 \\
    0 & 0 & 1 \\
  \end{pmatrix},\quad
M\=\frac{\omega}{2}
\begin{pmatrix}
    0 & -1 &0\\
1 & 0 & 0\\
0 & 0 & 0
  \end{pmatrix}
\]
give a 3-dimensional Lax representation for the harmonic oscillator.
\end{lemma}

\begin{defn}[quasi-canonical coordinates]
Define the  \emph{quasi-canonical coordinates} $A_\pm$ of HO by the relations
\begin{equation}
\label{eq:def_A}
A_+^2+A_-^2=2\sqrt{2H},\quad
A_+^2-A_-^2=2p,\quad
A_+A_-=\omega q
\end{equation}
Note that $A_\pm$ can not be simultaneously zero. 
\end{defn}

Denote by $\{\cdot,\cdot\}$ the ordinary \emph{Poisson brackets}. Then one has

\begin{thm}
\label{eq:quasi_poisson}
The quasi-canonical coordinates $A_{\pm}$ satisfy the relations
\begin{equation}
\label{eq:poisson_A}
\{A_+,A_+\}=0=\{A_-,A_-\},\quad \{A_+,A_-\}=\ee \=\frac{\omega}{2\sqrt{H}}
\end{equation}
\end{thm}

\begin{proof}
While the first two relations in \eqref{eq:poisson_A} are evident, we have only to check the third one. Calculate by using several times the Leibniz rule for the Poisson brackets:
\begin{align*}
2\omega \{p,q\}
&=\{A_+^2-A_-^2, A_+A_-\}\\
&=\{A_+^2, A_+A_-\}-\{A_-^2, A_+A_-\}\\
&=A_+\{A_+^2, A_-\}-\{A_-^2,A_+\}A_-\\
&=A_+\{A_+A_+, A_-\}-\{A_-A_-,A_+\}A_-\\
&=2(A_+^2+A_-^2)\{A_+,A_-\}\\
&=4\sqrt{2H}\{A_+,A_-\}
\end{align*}
Now use the relation $\{p,q\}=1$ to obtain the desired third formula in \eqref{eq:poisson_A}.
\end{proof}

\begin{thm}[\cite{PV08-1}]
\label{thm:main}
Let $C_{\nu}\in\mathbb{R}$ ($\nu=1,\ldots,9$) be
arbitrary real--valued parameters, such that
\begin{equation}
\label{eq:cond} C_2^2+C_3^2+C_5^2+C_6^2+C_7^2+C_8^2\neq0
\end{equation}
Let $M$ be defined as in Lemma \ref{lemma:harmonic3}, and
\begin{equation}\label{eq:theorem}
\begin{cases}
\mu_{11}^{1}=\mu_{22}^{1}=\mu_{33}^{1}=\mu_{11}^{2}=\mu_{22}^{2}=\mu_{33}^{2}=\mu_{11}^{3}=\mu_{22}^{3}=\mu_{33}^{3}=0\\
\mu_{23}^{1}=-\mu_{32}^{1}=C_2p-C_3\omega q-C_4\\
\mu_{13}^{2}=-\mu_{31}^{2}=C_2p-C_3\omega q+C_4\\
\mu_{31}^{1}=-\mu_{13}^{1}=C_2\omega q+C_3p-C_1\\
\mu_{23}^{2}=-\mu_{32}^{2}=C_2\omega q+C_3p+C_1\\
\mu_{12}^{1}=-\mu_{21}^{1}=C_5A_++C_6A_-\\
\mu_{12}^{2}=-\mu_{21}^{2}=C_5A_--C_6A_+\\
\mu_{13}^{3}=-\mu_{31}^{3}=C_7A_++C_8A_-\\
\mu_{23}^{3}=-\mu_{32}^{3}=C_7A_--C_8A_+\\
\mu_{12}^{3}=-\mu_{21}^{3}=C_9
\end{cases}
\end{equation}
Then $(\mu,M)$ is a $3$-dimensional anti-commutative binary operadic Lax pair of {\rm HO}.
\end{thm}

\section{Initial conditions}

Specify the coefficients $C_{\nu}$ in Theorem \ref{thm:main} by the
initial conditions
\[
\left. \mu\right|_{t=0}=\m{}_,\quad
\left.p\right|_{t=0}
=p_0,\quad \left. q\right|_{t=0}=0
\]
Denotinf $E\=H|_{t=0}$, the latter together with \eqref{eq:def_A} yield the initial
conditions for $A_{\pm}$:
\[
\begin{cases}
\left.\left(A_+^{2}+A_-^{2}\right)\right|_{t=0}=2\sqrt{2E}\\
\left.\left(A_+^{2}-A_-^{2}\right)\right|_{t=0}=2p_0\\
\left.A_+A_-\right|_{t=0}=0
\end{cases}
\quad \Longleftrightarrow \quad
\begin{cases}
p_0\!\!>0\\
\left.A^{2}_+\right|_{t=0}=2p_0\\
\left.A_-\right|_{t=0}=0
\end{cases}
\vee\quad
\begin{cases}
p_0<0\\
\left.A_+\right|_{t=0}=0\\
\left.A^2_-\right|_{t=0}=-2p_0
\end{cases}
\]
In what follows assume that $p_0>0$ and $A_+|_{t=0}=\sqrt{2p_0}$. The other cases
can be treated similarly. Note that in this case $p_0=\sqrt{2E}$. From \eqref{eq:theorem} we get the following linear system:
\begin{equation}
\label{eq:constants} \left\{
  \begin{array}{lll}
    \m{}_{23}^{1}=C_2p_0-C_4, & \m{}_{31}^{1}=C_3p_0-C_1, & \m{}_{12}^{1}=C_5\sqrt{2p_0}\\
    \m{}_{13}^{2}=C_2p_0+C_4, &
    \m{}_{12}^{2}=-C_6\sqrt{2p_0}, &
    \m{}_{23}^{2}=C_3p_0+C_1\\
    \m{}_{13}^{3}=C_7\sqrt{2p_0}, &
\m{}_{23}^{3}=-C_8\sqrt{2p_0}, & \m{}_{12}^{3}=C_9
\end{array}
\right.
\end{equation}
One can easily check that the unique solution of the latter system
with respect to $C_\nu$ ($\nu=1,\ldots,9$) is
\[
\left\{
  \begin{array}{lll}
C_1=\frac{1}{2}\left(\overset{\circ}{\mu}{}_{23}^{2}-\overset{\circ}{\mu}{}_{31}^{1}\right),&
C_2=\frac{1}{2p_0}\left(\overset{\circ}{\mu}{}_{13}^{2}+\overset{\circ}{\mu}{}_{23}^{1}\right),&
C_3=\frac{1}{2p_0}\left(\overset{\circ}{\mu}{}_{23}^{2}+\overset{\circ}{\mu}{}_{31}^{1}\right)\vspace{1mm}\\
C_4=\frac{1}{2}\left(\overset{\circ}{\mu}{}_{13}^{2}-\overset{\circ}{\mu}{}_{23}^{1}\right),&
C_5=\frac{1}{\sqrt{2p_0}}\overset{\circ}{\mu}{}_{12}^{1},&
C_6=-\frac{1}{\sqrt{2p_0}}\overset{\circ}{\mu}{}_{12}^{2}\vspace{1mm}\\
C_7=\frac{1}{\sqrt{2p_0}}\overset{\circ}{\mu}{}_{13}^{3},&
C_8=-\frac{1}{\sqrt{2p_0}}\overset{\circ}{\mu}{}_{23}^{3},&
C_9=\overset{\circ}{\mu}{}_{12}^{3}
\end{array}
\right.
\]

\section{VII$_{a}$, III$_{a=1}$, VI$_{a\neq1}$}

We study only the algebras {\rm VII}$_{a}$, {\rm III}$_{a=1}$, {\rm VI}$_{a\neq1}$ from the Bianchi classification of the 3-dimensional real Lie algebras \cite{Landau80}.
The structure equations of the 3-dimensional real Lie algebras can be presented
as follows:
\[
[e_1,e_2]=-\alpha e_2+n^{3}e_3,\quad
[e_2,e_3]=n^{1}e_1,\quad
[e_3,e_1]=n^{2}e_2+\alpha e_3
\]
The values of the parameters $\alpha,n^{1}, n^{2},n^{3}$  and the corresponding structure constants for {\rm VII}$_{a}$, {\rm III}$_{a=1}$, {\rm VI}$_{a\neq1}$  are presented in Table \ref{table:Bianchi1}. 
\begin{table}[ht]
\begin{center}
\begin{tabular}{|c||c||c|c|c||c|c|c|c|c|c|c|c|c|c|c|}\hline
Bianchi type & $\alpha$ & $n^{1}$ & $n^{2}$ & $n^{3}$ &
$\overset{\circ}{\mu}{}_{12}^{1}$ &
$\overset{\circ}{\mu}{}_{12}^{2}$ &
$\overset{\circ}{\mu}{}_{12}^{3}$ &
 $\overset{\circ}{\mu}{}_{23}^{1}$ & $\overset{\circ}{\mu}{}_{23}^{2}$ & $\overset{\circ}{\mu}{}_{23}^{3}$
  & $\overset{\circ}{\mu}{}_{31}^{1}$ & $\overset{\circ}{\mu}{}_{31}^{2}$ &
  $\overset{\circ}{\mu}{}_{31}^{3}$\\\hline\hline
{\rm VII}$_{a}$& a & 0 & $1$ & $1$ & 0 & $-a$ & $-1$ & 0 & 0 & 0 & 0 &
$1$ & $a$ \\\hline
{\rm III}$_{a=1}$& 1 & 0 & $1$ & $-1$ & 0 & $-1$ & $-1$ & 0 & 0 & 0 & 0 &
$1$ & $1$
\\\hline
{\rm VI}$_{a\neq 1}$& $a$ & 0 & $1$ & $-1$ & 0 & $-a$ & $-1$ & 0 & 0 & 0 &
0 & $1$ & $a$
\\\hline
\end{tabular}
\end{center}
\caption{{\rm VII}$_{a}$, {\rm III}$_{a=1}$, {\rm VI}$_{a\neq1}$. Here $a>0$.}
\label{table:Bianchi1}
\end{table}

\section{VII$_{a}^{t}$, III$_{a=1}^{t}$, VI$_{a\neq1}^{t}$}

By using the structure constants of the 3-dimensional Lie algebras
in the Bianchi classification, Theorem \ref{thm:main} and relations
\eqref{eq:constants} one can propose that evolution of the {\rm VII}$_{a}$, {\rm III}$_{a=1}$, {\rm VI}$_{a\neq1}$ can be prescribed as given in Table \ref{table:Bianchi3}.

\begin{table}[!h]
\begin{center}\setlength\extrarowheight{4pt}
\begin{tabular}{|c||c|c|c|c|c|c|c|c|c|c|c|}\hline
Dynamical Bianchi type & $\mu_{12}^{1}$ & $\mu_{12}^{2}$ &
$\mu_{12}^{3}$ & $\mu_{23}^{1}$ & $\mu_{23}^{2}$ & $\mu_{23}^{3}$ &
$\mu_{31}^{1}$ & $\mu_{31}^{2}$ &  $\mu_{31}^{3}$
\\[1.5ex]\hline\hline
{\rm VII}$^{t}_a$ & $\frac{aA_-}{\sqrt{2p_0}}$ &
$\frac{-aA_+}{\sqrt{2p_0}}$ & $1$ & $\frac{p-p_0}{-2p_0}$ &
$\frac{\omega q}{-2p_0}$ & $\frac{-aA_-}{\sqrt{2p_0}}$ &
$\frac{\omega q}{-2p_0}$ & $\frac{p+p_0}{2p_0}$ &
$\frac{aA_+}{\sqrt{2p_0}}$
\\ [1.5ex] \hline
{\rm III}$_{a=1}^{t}$ & $\frac{A_-}{\sqrt{2p_0}}$ &
$\frac{-A_+}{\sqrt{2p_0}}$ & $-1$ & $\frac{p-p_0}{-2p_0}$ &
$\frac{\omega q}{-2p_0}$ & $\frac{-A_-}{\sqrt{2p_0}}$ & $\frac{\omega
q}{-2p_0}$ & $\frac{p+p_0}{2p_0}$ & $\frac{A_+}{\sqrt{2p_0}}$
\\ [1.5ex] \hline
{\rm VI}$_{a\neq1}^{t}$ & $\frac{aA_-}{\sqrt{2p_0}}$ &
$\frac{-aA_+}{\sqrt{2p_0}}$ & $-1$ & $\frac{p-p_0}{-2p_0}$ &
$\frac{\omega q}{-2p_0}$ & $\frac{-aA_-}{\sqrt{2p_0}}$ &
$\frac{\omega q}{-2p_0}$ & $\frac{p+p_0}{2p_0}$ &
$\frac{aA_+}{\sqrt{2p_0}}$
\\ [1.5ex] \hline
\end{tabular}
\end{center}
\caption{{\rm VII}$_{a}^{t}$, {\rm III}$_{a=1}^{t}$, {\rm VI}$_{a\neq1}^{t}$}
\label{table:Bianchi3}
\end{table}

\section{Quantum conditions}

In the dynamically deformed algebras {\rm III}$_{a=1}^{t}$  from Table \ref{table:Bianchi3}
the structure functions $\mu^{i}_{jk}$ depend on the canonical and quasi-canonical coordinates of the harmonic oscillator. The quasi-canonical coordinates $A_\pm$ of HO were defined by relations \eqref{eq:def_A}.  Following the canonical quantization prescription, the corresponding quantum  relations for  $\A_{\pm}$ read

\begin{equation}
\label{eq:qdef_A}
\A_+^2+\A_-^2=2\sqrt{2\H},\quad
\A_+^2-\A_-^2=2\pp,\quad
\A_+\A_-+\A_-\A_+=2\omega \q
\end{equation}
In what follows, we shall use the Schr\"odinger picture, i.e the operators 
$\q,\pp,\H,\A_\pm$ do not depend on time.
Denote by $[\cdot,\cdot]$ the ordinary commutator bracketing.
The quantum counterpart of \eqref{eq:quasi_poisson}, called the \emph{quasi-canonical} commutation relations (quasi-CCR)  read
\begin{equation}
\label{eq:qpoisson_A}
[\A_+,\A_+]=0=[\A_-,\A_-],\quad 
[\A_+,\A_-]=\frac{\hbar}{i}\e\=\frac{\hbar}{i}\frac{\omega}{2\sqrt{2\H}}
\end{equation}

\section{VII$_{a}^{\hbar}$, III$_{a=1}^{\hbar}$, VI$_{a\neq1}^{\hbar}$}

By using the algebras {\rm VII}$_{a}^{t}$, {\rm III}$_{a=1}^{t}$, {\rm VI}$_{a\neq1}^{t}$
from Table \ref{table:Bianchi3}, we can now propose their quantized counterparts 
{\rm VII}$_{a}^{\hbar}$, {\rm III}$_{a=1}^{\hbar}$, {\rm VI}$_{a\neq1}^{\hbar}$ as exposed in Table \ref{table:Bianchi4}. We shall fix the parameter $p_0>0$ later.

\begin{table}[!h]
\begin{center}\setlength\extrarowheight{4pt}
\begin{tabular}{|c||c|c|c|c|c|c|c|c|c|c|c|}\hline
Quantum Bianchi type & $\muu_{12}^{1}$ & $\muu_{12}^{2}$ &
$\muu_{12}^{3}$ & $\muu_{23}^{1}$ & $\muu_{23}^{2}$ &
$\muu_{23}^{3}$ & $\muu_{31}^{1}$ & $\muu_{31}^{2}$ &
$\muu_{31}^{3}$
\\[1.5ex]\hline\hline
{\rm VII}$^{\hbar}_a$ & $\frac{a\A_-}{\sqrt{2p_0}}$ &
$\frac{-a\A_+}{\sqrt{2p_0}}$ & $1$ & $\frac{\pp-p_0}{-2p_0}$ &
$\frac{\omega \q}{-2p_0}$ & $\frac{-a\A_-}{\sqrt{2p_0}}$ &
$\frac{\omega \q}{-2p_0}$ & $\frac{\pp+p_0}{2p_0}$ &
$\frac{a\A_+}{\sqrt{2p_0}}$
\\ [1.5ex] \hline
{\rm III}$_{a=1}^{\hbar}$ & $\frac{\A_-}{\sqrt{2p_0}}$ &
$\frac{-\A_+}{\sqrt{2p_0}}$ & $-1$ & $\frac{\pp-p_0}{-2p_0}$ &
$\frac{\omega \q}{-2p_0}$ & $\frac{-\A_-}{\sqrt{2p_0}}$ & $\frac{\omega
\q}{-2p_0}$ & $\frac{\pp+p_0}{2p_0}$ & $\frac{\A_+}{\sqrt{2p_0}}$
\\ [1.5ex] \hline
{\rm VI}$_{a\neq1}^{\hbar}$ & $\frac{a\A_-}{\sqrt{2p_0}}$ &
$\frac{-a\A_+}{\sqrt{2p_0}}$ & $-1$ & $\frac{\pp-p_0}{-2p_0}$ &
$\frac{\omega \q}{-2p_0}$ & $\frac{-a\A_-}{\sqrt{2p_0}}$ &
$\frac{\omega \q}{-2p_0}$ & $\frac{\pp+p_0}{2p_0}$ &
$\frac{a\A_+}{\sqrt{2p_0}}$
\\ [1.5ex] \hline
\end{tabular}
\end{center}
\caption{{\rm VII}$_{a}^{\hbar}$, {\rm III}$_{a=1}^{\hbar}$, {\rm VI}$_{a\neq1}^{\hbar}$}
\label{table:Bianchi4}
\end{table}
Denoting $\muu:=[\cdot,\cdot]_\hbar$, the quantum Jacobian is defined by
\begin{align*}
\hat{J}_\hbar(x;y;z)
&:=[[x,y]_\hbar,z]_\hbar+[[y,z]_\hbar,x]_\hbar+[[z,x]_\hbar,y]_\hbar\\
&\,\,=\hat{J}^1_\hbar(x,y,z)e_1+\hat{J}^2_\hbar(x,y,z)e_2+\hat{J}^3_\hbar(x,y,z)e_3
\end{align*}
In \cite{PV09-1} we calculated the Jacobians for all 3d real Lie algebras from the Bianchi classification. Here we concentrate only on {\rm III}$_{a=1}^{\hbar}$, {\rm VI}$_{a\neq1}^{\hbar}$, and {\rm VII}$^{\hbar}_a$ . 
Denote the volume of a triple $x,y,z\in\RR^3$ by
\begin{equation*}
(x,y,z)\= 
\begin{vmatrix}
 x^{1} & x^{2} & x^{3} \\
 y^{1} & y^{2} & y^{3} \\
z^{1} & z^{2} & z^{3} \\
\end{vmatrix},\quad
\qxi_{\pm} \=\omega\q\A_\mp\pm (\pp \mp p_0)\A_\pm
\end{equation*}
Recall
\begin{thm}[\cite{PV09-1,PV11}]
The Jacobian coordinates of {\rm VI}$_{a\neq1}^{\hbar}$ and {\rm VII}$^{\hbar}_a$ read
\[
      \hat{J}^{1}_\hbar(x;y;z)=\frac{a\tau(x,y,z)}{\sqrt{2p_0^{3}}}\qxi_{+},\quad
      \hat{J}^{2}_\hbar(x;y;z)=\frac{a\tau(x,y,z)}{\sqrt{2p_0^{3}}}\qxi_{-},\quad
      \hat{J}^{3}_\hbar(x;y;z)=\frac{a^{2}(x,y,z)}{p_0}[\A_+,\A_-]
 \]
with $\tau=1$ for {\rm VI}$_{a\neq1}^{\hbar}$ and $\tau=-1$ for {\rm VII}$_{a\neq1}^{\hbar}$. For {\rm III}$_{a=1}^{\hbar}$ one has the same formulae with $a=\tau=1$.
\end{thm}

\begin{lemma}
Let $p_0\=\sqrt{2E}$. We have
\begin{equation*}
\qxi_{\pm}=A_\pm\left(\sqrt{2\H} - \sqrt{2E}\right)\pm
\frac{\hbar}{i}\A_\mp\frac{\e}{2}
\end{equation*}
\end{lemma}

\begin{proof}

By using relations \eqref{eq:qdef_A} and \eqref{eq:qpoisson_A} calculate:
\begin{align*}
\qxi_{+}
&=\omega\q\A_- + (\pp - p_0)\A_+\\
&=\frac{1}{2}(\A_+\A_-+\A_-\A_+)\A_- +\frac{1}{2}(\A^{2}_+-\A^{2}_-)\A_+ - p_0 \A_+\\
&=\frac{1}{2}(\A_+\A^{2}_-+\A_-\A_+\A_- +\A^{3}_+ - \A^{2}_-\A_+)  -p_0\A_+\\
&=\frac{1}{2}\left[\A_-(\A_+\A_- - \A_-\A_+)+\A_+(\A^{2}_- + \A^{2}_+)\right]-p_0\A_+\\
&=\frac{1}{2}\A_-[\A_+,\A_-]+\frac{1}{2}\A_+(\A^{2}_+ + \A^{2}_-)  -p_0\A_+\\
&=\frac{\hbar}{i}\A_-\frac{\e}{2} +A_+\sqrt{2\H} - \sqrt{2E}A_+\\
&=\frac{\hbar}{i}\A_-\frac{\e}{2} +A_+\left(\sqrt{2\H} - \sqrt{2E}\right)
 \end{align*}
 In the same way one can calculate $\qxi_{-}$.
Thus,
\begin{align*}
\label{eq:J}
\hat{J}^{1}_\hbar(x;y;z)
&=\frac{a\tau(x,y,z)}{\sqrt{(2p_0)^3}} 
\left[A_+\left(\sqrt{2\H} - \sqrt{2E}\right)+\frac{\hbar}{i}\A_- \e \right]\\
\hat{J}^{2}_\hbar(x;y;z)
&= \frac{a\tau(x,y,z)}{\sqrt{(2p_0)^3}} 
\left[ A_-\left(\sqrt{2\H} - \sqrt{2E}\right) - \frac{\hbar}{i}\A_+ \e \right]\\
\hat{J}^{3}_\hbar(x;y;z)
&=\frac{\hbar}{i}  \frac{a^{2}(x,y,z)}{p_0} \e
\tag*{\qed}
\end{align*}
\renewcommand{\qed}{}
\end{proof}

\section{Derivative algebra of VII$^{\hbar}_a$,  III$_{a=1}^{\hbar}$, VI$_{a\neq1}^{\hbar}$}

\begin{lemma}
The Jacobi operator coordinates of {\rm VII}$^{\hbar}_a$, {\rm III}$_{a=1}^{\hbar}$, {\rm VI}$_{a\neq1}^{\hbar}$, 
 generate a 3-dimensional real anti-commutative algebra with sturucture equations
\begin{equation}
\label{eq:TA}
[\hat{J}^{1}_\hbar,\hat{J}^{3}_\hbar]=0=[\hat{J}^{2}_\hbar,\hat{J}^{3}_\hbar], \quad
[\hat{J}^{1}_\hbar,\hat{J}^{2}_\hbar]=C_n\hat{J}^{3}_\hbar
\end{equation}
where
\[
C_n\=\left(\frac{\hbar\omega}{2E_n}\right)^{2}\frac{(x,y,z)}{32} , \quad 
E_n\=\hbar\omega\left(n+\frac{1}{2}\right),
\quad n=0,1,2\ldots
\]
\end{lemma}

\begin{proof}
In the state space of the harmonic oscillator use the basis of the energy eigenvectors, i.e $\H\ket{n}=E_n\ket{n}$.
Now fix in the Jacobi operator coordinates the value of the free parameter $E\=p^{2}_0/2$ to be $E\=E_n$ and  denote
\[
\ee_n\=\frac{\omega}{2\sqrt{2E_n}}, \quad
 \la_n \=-\frac{\hbar}{i}\frac{a\tau(x,y,z)}{\sqrt{(2\pn)^{3}}} \ee_n
\]
Calculate the value of the Jacobi operator coordinates on the eigenvector $\ket{n}$:
\begin{align*}
\hat{J}^{1}_\hbar(x,y,z) \ket{n}
&=\frac{a\tau(x,y,z)}{\sqrt{(2\pn)^3}} 
\left[A_+\left(\sqrt{2\H} - \sqrt{2E}\right)+\frac{\hbar}{i}\A_- \e \right]\ket{n} \\
&=\frac{a\tau(x,y,z)}{\sqrt{(2\pn)^3}}\frac{\hbar}{i}\A_-  \ee_n \ket{n}\\
&=-\la_n \A_- \ket{n}\\
 \hat{J}^{2}_\hbar(x,y,z) \ket{n}
&= \frac{a\tau(x,y,z)}{\sqrt{(2\pn)^3}} 
\left[ A_-\left(\sqrt{2\H} - \sqrt{2E}\right) - \frac{\hbar}{i}\A_+ \e \right]  \ket{n}\\
&=-\frac{a\tau(x,y,z)}{\sqrt{(2\pn)^3}}\frac{\hbar}{i}\A_+  \ee_n \ket{n}\\
&=\la_n A_+\ket{n}\\
\hat{J}^{3}_\hbar(x,y,z)  \ket{n}
&=\frac{\hbar}{i}  \frac{a^{2}(x,y,z)}{\pn} \ee_n  \ket{n}
\end{align*}
As soon as $\ket{n}$ ($n=0,1,2,\ldots$) form a basis we have
\[
\hat{J}^{1}_\hbar=-\la_n \A_-,\quad  \hat{J}^{2}_\hbar=\la_n \A_+
\]
Calculate:
\begin{align*}
[\hat{J}^{1}_\hbar,\hat{J}^{2}_\hbar]   \ket{n} 
&=-\la^{2}_n[\A_+,\A_-]    \ket{n}  \\
&=-\la^{2}_n\frac{\hbar}{i} \e    \ket{n} \\
&=-\la^{2}_n\frac{\hbar}{i} \ee_n     \ket{n}\\
&=-\left(\frac{\hbar}{i} \ee_n\right)^{2}   
\frac{a^{2}(x,y,z)}{2\pn} \frac{(x,y,z)}{(2\pn)^{2}} \frac{\hbar}{i} \ee_n  \ket{n} \\
&= - \left(\frac{\hbar}{i} \ee_n \right)^{2}  \frac{(x,y,z)}{2(2\pn)^{2}} 
\hat{J}^{3}_\hbar   \ket{n} \\
&= - \left(\frac{\hbar}{i} \frac{\omega}{2\sqrt{2E_n}} \right)^{2}  \frac{(x,y,z)}{8\cdot 2E_n} \hat{J}^{3}_\hbar   \ket{n}  \\
&=  \left(\frac{\hbar\omega}{2E_n}\right)^{2}\frac{(x,y,z)}{32}\hat{J}^{3}_\hbar  \ket{n}\\
&=C_n \hat{J}^{3}_\hbar   \ket{n}
\tag*{\qed}
\end{align*}
\renewcommand{\qed}{}
\end{proof}

\begin{defn}
[derivative algebra of {\rm VII}$^{\hbar}_a$,  {\rm III}$_{a=1}^{\hbar}$, {\rm VI}$_{a\neq1}^{\hbar}$]
The anti-commutative algebra given by the structure equations \eqref{eq:TA} is called the \emph{derivative algebra} of {\rm III}$_{a=1}^{\hbar}$, {\rm VI}$_{a\neq1}^{\hbar}$, {\rm VII}$^{\hbar}_a$.
\end{defn}

\begin{cor}
Define the new basis in the derivative algebra:
\[
e_1=(x,y,z)\hat{J}^{3}_\hbar, \quad
e_2=(x,y,z)\hat{J}^{1}_\hbar, \quad
e_3=(x,y,z)\hat{J}^{2}_\hbar
\]
Then the structure equations of the derivative algebra read
\[
\quad [e_1,e_3]=0=[e_2,e_3], \quad [e_2,e_3]=\beta^{2}_ne_1\]
where
\[
\beta_n= \frac{\hbar\omega}{2E_n} \frac{|(x,y,z)|}{4\sqrt{2}} > 0
\]
\end{cor}

\begin{proof}
Calculate:
\begin{equation*}
[e_2,e_3]
=(x,y,z)(x,y,z)[\hat{J}^{1}_\hbar,\hat{J}^{2}_\hbar]
=C_n(x,y,z)(x,y,z)\hat{J}^{3}_\hbar
=C_n(x,y,z)e_1
=\beta^{2}_n e_1 
\tag*{\qed}
\end{equation*}
\renewcommand{\qed}{}
\end{proof}

\begin{thm}
The derivative algebra of {\rm VII}$^{\hbar}_a$, {\rm III}$_{a=1}^{\hbar}$, {\rm VI}$_{a\neq1}^{\hbar}$ is the 3-dimensional real Heisenberg algebra.
\end{thm}

\begin{proof}
By elementary calculus one can see that the Jacobi operator of the derivative algebra vanish. As the only non-vanishing structure constant is $\overset{\circ}{\mu}{}_{23}^{1}$, one can easily see from the Bianchi  classification \cite{Landau80} that $\beta_n=1$ perfectly suits.
\end{proof}

\begin{cor}
In  {\rm III}$_{a=1}^{\hbar}$, {\rm VI}$_{a\neq1}^{\hbar}$, {\rm VII}$^{\hbar}_a$, the volume is quantized by $|(x,y,z)|=4\sqrt{2}(2n+1)$, ($n=0,1,2,\dots$). Thus, the elementary (minimal) length in this model is $l_{min}=2^{5/6}$.
\end{cor}

The research was in part supported by the Estonian Research Council, Grants ETF-6912 and ETF-9038. Authors are grateful to S. Hervik and P. Kuusk for discussions about the Bianchi cosmologies.

\noindent
{\footnotesize
Tallinn University of Technology,
Ehitajate tee 5, 19086 Tallinn, Estonia
}

\end{document}